\documentclass[11pt]{article}
\usepackage{graphicx}
\usepackage{amsmath}
\usepackage{amsfonts}
\usepackage{amssymb}
\usepackage{epsfig}
\usepackage{times}
\usepackage{cite}
\usepackage{calc}
\usepackage{version}
\usepackage[french,english]{babel}
\begin{document}


\begin{titlepage}


\null

\vskip 2.5cm

{\bf\large\baselineskip 20pt
\begin{center}
\begin{Large}
Medium-modified evolution of multiparticle production in jets in
heavy-ion collisions
\end{Large}
\end{center}
}
\vskip 1cm

\begin{center}

Redamy P\'erez-Ramos\footnote{E-mail: redamy@mail.desy.de}\\
\smallskip
II. Institut f\"ur Theoretische Physik, Universit\"at Hamburg\\
Luruper Chaussee 149, D-22761 Hamburg, Germany
\end{center}

\baselineskip=15pt

\vskip 3.5cm

{\bf Abstract}: The energy evolution of medium-modified 
average multiplicities and multiplicity fluctuations in quark and 
gluon jets produced in heavy-ion collisions is 
investigated from 
a toy QCD-inspired model. 
In this model, we use modified splitting functions 
accounting for medium-enhanced radiation of gluons 
by a fast parton which propagates through the
quark gluon plasma. The leading contribution of the standard production of soft
hadrons is found to be enhanced by the factor $\sqrt{N_s}$ while
next-to-leading order (NLO) corrections are suppressed by $1/\sqrt{N_s}$, where
the nuclear parameter $N_s>1$ accounts for the 
induced-soft gluons in the hot medium. The role
of next-to-next-to-leading order corrections (NNLO) is studied
and the large amount of medium-induced soft gluons is found to drastically 
affect the convergence of the perturbative series.
Our results for such global observables are 
cross-checked and compared 
with their limits in the vacuum and a new method for solving the 
second multiplicity correlator evolution equations
is proposed.

\vskip .5 cm


\vfill


\end{titlepage}

\section{Introduction}
The properties of quark and gluon jets are strongly established and carefully
studied through well known QCD evolution equations \cite{Basics}
in $e^+e^-$, $ep$ and hadron collision experiments. In the case of
high energy nucleus-nucleus collisions, hard jets propagate in a 
medium with different properties from those of the vacuum.
Recent experiments at the Relativistic Heavy Ion Collider (RHIC) have established
a phenomenon of strong high-transverse momentum hadron suppression 
\cite{PHENIX,STAR}, which supports the picture 
that hard partons going through dense matter
suffer a significant energy loss prior to hadronization
in the vacuum (for recent review see \cite{arleo}). 

Since little is known so far on jet evolution in QCD media,
predictions for multiparticle production
in such reactions could be carried out by using a toy QCD-inspired model 
introduced by Borghini and Wiedemann in \cite{Urs};
it allows for analytical computations and may
capture some important features of a more complete QCD description. 
In this model, the Dokshitzer-Gribov-Lipatov-Altarelli-Parisi 
(DGLAP) splitting functions $q\to g\bar q$ and $g\to gg$ \cite{Basics}
of the QCD evolution equations 
were distorted so that the role of soft emissions was
enhanced by multiplying the infra-red
singular terms by the medium factor $N_s$:
$$
\Phi_g^g(N_s,x)=\frac{N_s}{x}-(1-x)[2-x(1-x)],\quad
\Phi_q^g(N_s,x)=\frac{C_F}{N_c}\left(\frac{N_s}{x}-1+\frac{x}2\right),
$$
where $x$ is the fraction of the outgoing jet energy carried away by
a single gluon. Thus, the leading singular terms of the splitting functions play
a more important role; from the theoretical point of view it could be considered
as a result of some effective Lagrangian, which would be responsible for processes 
in a dense nuclear environment. 
In the approach \cite{leticia}, the medium-modified splitting functions
are directly related to the medium-induced gluon spectrum $dI^{\text{med}}/dxdE$
\cite{wiedemann}, where $E$ is the initial energy of the emitting
parton going through the medium. 
As compared to the Borghini-Wiedemann model, the medium
modifications explicitly depend on the parton virtuality through
the enhanced induced gluon spectrum. However, we use the
simpler interpretation of the induced-medium 
modification of the Borghini-Wiedemann model \cite{Urs}, which 
was further discussed and used on the 
description of final state hadrons produced in heavy-ion 
collisions \cite{sapeta}. 

In this paper we are concerned with multiparticle production 
in quark and gluon jets $A=q,g$, produced
in nucleus-nucleus collisions at very high energy. We make predictions 
at NLO and NNLO for the average multiplicities $N_A$, 
for the ratio $r=N_g/N_q$ 
and finally for the second multiplicity correlators 
$\langle N_A(N_A-1)\rangle/N_A^2$,
which defines the width of the multiplicity distribution.
Such observables are of great importance and have been 
largely studied in the vacuum from both theoretical 
\cite{MultTheory,MALAZA,DREMIN,dln} and 
experimental \cite{OPALrgq,DELPHIrgq,experN2,DreminGary,TevMult,TevRap} 
points of view. The problem of medium-modified multiparticle 
production has also been considered in \cite{termo,paloma}
with fixed coupling constant.

The starting point of our analysis is the NLO or
Modified-Leading-Logarithmic-Approximation (MLLA)
evolution equations \cite{Basics}, which determine the jet properties
at all energies together with the initial conditions at 
threshold at small $x$.
Their solutions with medium-modified splitting functions
can be resummed in powers of
$\sqrt{\alpha_s/N_s}$ and the leading contribution can be 
represented as an exponential of the medium-modified anomalous 
dimension, which takes the $N_s$-dependence:
$$
N_A\simeq \exp\left\{\int^{Y}
\gamma_{\text{med}}\left(\alpha_s(Y)\right)dY\right\},
$$
where $\gamma_{\text{med}}(\alpha_s)$ can be expressed as
a power series of $\sqrt{\alpha_s/N_s}$ in the symbolic form:
$$
\gamma_{\text{med}}\left(\alpha_s\right)
\simeq\sqrt{N_s}\times\sqrt{\alpha_s}\left(1+\sqrt{\frac{\alpha_s}{N_s}}+
\frac{\alpha_s}{N_s}+\ldots\right).
$$
Within this logic, the leading double logarithmic approximation 
(DLA, ${\cal O}(\sqrt{N_s\alpha_s})$), which resums both
soft and collinear gluons, and NLO 
(MLLA, ${\cal O}(\alpha_s)$), which resums hard collinear partons
and accounts for the running of the coupling constant $\alpha_s$,
are complete. The DLA 
takes into account, as expected, the medium modification by enhancing
the soft multiparticle production by a factor 
$\propto\sqrt{N_s}$, the MLLA terms, which are $N_s$-independent, 
takes into account the energy balance in the hard collinear parton 
splitting region as in the absence of the nuclear modification. However,
this result depends on the definition of the medium-modified splitting
functions. The next terms, 
which are NNLO or next-to-MLLA (NMLLA, ${\cal O}(\alpha_s^{3/2}/\sqrt{N_s})$) are not
complete but they include an important contribution, which takes into
account energy conservation and provide an improved behavior near threshold.
With medium modification, the NMLLA terms take $N_s$-dependence, but 
this will be explained in the main core of the paper. This logic applies to each
vertex of the cascade and the solution represents the fact that successive and
independent partonic splittings inside the shower, 
which in this case concern both vacuum and medium-induced soft gluons,
exponentiate with respect to the {\em evolution-time} variable Y 
($dY=d\Theta/\Theta$), where $\Theta\ll1$ is the angle between outgoing
couples of partons. The choice of $Y\simeq\ln(\Theta)$ follows from Angular
Ordering (AO) in intra-jet cascades; furthermore, the tree Feynman 
diagrams describing the process are at the heart of the 
{\em parton shower picture} \cite{Basics}. Thus, the solutions of the 
equations incorporate the Markov chains of sequential angular ordered
decays and $\gamma_{\text{med}}$ determines, in this case, the rate
of multiparticle production in the dense medium. 

At the end of the cascading process inside the medium, partons
hadronize in the vacuum. In order to obtain the hadronic spectra,
we advocate for the Local Parton Hadron Duality (LPHD) hypothesis \cite{LPHD}:
global and differential partonic observables can be normalized 
to the corresponding charged hadronic observables via a certain constant 
${\cal K}$ that can be fitted to the data, i.e. $N^h_{A}={\cal K}\times N_{A}$.

The paper is organized as follows:

\begin{itemize}
\item{Section \ref{section:navemult} presents 
a system of evolution equations
with medium-modified splitting functions,
which allows for the computation of the medium-modified average
multiplicity and the medium-modified gluon to quark average multiplicity ratio
at NLO and NNLO. We give predictions for the values $N_s=1.6$
and $N_s=1.8$, which may be realistic for RHIC and LHC phenomenology
\cite{Urs}. Moreover, we compare our
results with previous predictions in the vacuum;}

\item{in Section \ref{section:2corr} we study the medium-modified
second multiplicity correlator
at NLO and NNLO. Accordingly, 
we give predictions for the same values of $N_s$ and compare such
predictions with the equivalent for the vacuum limit $N_s=1$;}

\item{in Section \ref{section:conclusions} we present our conclusions.}
\end{itemize}

\section{Evolution of the average multiplicity
and gluon to quark average multiplicity ratio with energy loss}

\label{section:navemult}

At MLLA the evolution of quantities with jet energy $E$
and jet opening angle $\Theta$ is given by an evolution equation for the
azimuthally averaged generating functional in 
the jet \cite{Basics}. The evolution involves 
$\alpha_s$, the running coupling constant of QCD:
\begin{equation}\label{eq:varbis}
\alpha_s\equiv\alpha_s(Q)=\frac{2\pi}{4N_c\beta_0
\ln\left(\frac{Q}{\Lambda}\right)},\quad
\beta_0=\frac1{4N_c}\left(\frac{11}3N_c-\frac43T_R\right),
\end{equation}
where $Q=E\Theta$ is the maximum transverse momentum of the jet,
$\Lambda\equiv\Lambda_{QCD}$ is the intrinsic scale of
QCD, $\beta_0$ is the first term in the perturbative expansion
of the $\beta-$function, $N_c$ is the number of colors and 
$T_R=n_f/2$, where $n_f$ is the number of quark flavors.
In the leading DLA, $\alpha_s$ is
linked to the anomalous dimension $\gamma_0$ of twist-2
operators by the formula:
\begin{equation}\label{eq:anodim}
\gamma_0^2\equiv\gamma_0^2(Q) = 2N_c\frac{\alpha_s(Q)}{\pi}=
\frac1{\beta_0(Y+\lambda)},\quad
Y=\ln\frac{Q}{Q_0},\quad\lambda
=\ln\frac{Q_0}{\Lambda},
\end{equation}
where $Q_0$ is the collinear cut-off parameter $k_T=E\Theta>Q_0$. 
The results depend on energy and
angle only through the variable $Y$.
For the sake of simplicity
we also set $Y'=Y+\lambda$ in the following. The average multiplicity
is obtained by integrating the one-particle
single differential inclusive cross section over the energy fraction
$x=e/E$
$$
N_A(Y)=\int dx\left(\frac{1}{\sigma}\frac{d\sigma}{dx}\right).
$$
For the medium-modified evolution of the average
multiplicity in quark and gluon jets 
one obtains as a consequence of AO at MLLA,
the coupled system of two evolution equations 
\begin{eqnarray}
\frac{d}{dY}N_g(Y)\!\!&\!\!=\!\!&\!\!
\!\!\int_0^1 dx\,\gamma_0^2(x)\left[\Phi_g^g(N_s,x)\left(N_g(Y+\ln x)+
N_g(Y+\ln(1-x))-N_g(Y)\right)\notag\right.\\
\!\!&\!\!\!\!&\!\!\left.\hskip 0.5cm+n_f\Phi_g^q(N_s,x)
\left(N_q(Y+\ln x)+
N_q(Y+\ln(1-x))-N_g(Y)\right)\right],\label{eq:NGh}\\\notag\\
\frac{d}{dY}N_q(Y)\!\!&\!\!=\!\!&\!\!\!\!\int_0^1 dx\,
\gamma_0^2(x)\!\!\left[\Phi_q^g(N_s,x)\left(N_g(Y+\ln x)+
N_q(Y+\ln(1-x))-N_q(Y)\right)\right]
\label{eq:NQh}
\end{eqnarray}
with medium-modified splitting functions as suggested in \cite{Urs}
in the Borghini-Wiedemann model

\begin{equation}\label{eq:splitG}
\Phi_g^g(N_s,x)=\frac{N_s}{x}-(1-x)[2-x(1-x)],\quad
\Phi_g^q(N_s,x)=\frac1{4N_c}[x^2+(1-x)^2],
\end{equation}
\begin{equation}\label{eq:splitQ}
\Phi_q^g(N_s,x)=\frac{C_F}{N_c}\left(\frac{N_s}{x}-1+\frac{x}2\right),
\end{equation}
which accounts for energy loss in the medium by enhancing the singular terms
like $\Phi\approx N_s/x$ as $x\to0$. 
The $g\to q\bar q$ splitting function as well as the regular
parts of $g\to gg$ and $q\to g\bar q$ splitting functions
are hard and provide collinear corrections, that is why these terms do 
not take $N_s$ dependence.

\subsection{MLLA evolution of the average multiplicity in the medium}

\label{sub:eqmlla}

For $Y\gg\ln x\sim\ln(1-x)$, 
$N(Y+\ln x)$ ($N(Y+\ln(1-x))$) can be
replaced by $N(Y)$ in the hard partonic splitting region
$x\sim1$ ($1-x\sim1$) (non-singular or regular parts of the splitting functions), 
while  the dependence on $\ln x$ is kept on the singular
one $\Phi(x)\approx N_s/x$ as it is performed in the vacuum. Furthermore, the integration
over $x$ can be replaced by the integration over 
$Y(x)=\ln\left(\frac{xE\Theta}{Q_0}\right)$

\begin{equation}
\int^1\gamma_0^2(x)\frac{dx}x=\int^{Y}\gamma_0^2(Y(x))dY(x).
\end{equation}
After applying the differential operator 
$\frac{d}{dY}$ to both members of the system (\ref{eq:NGh},\ref{eq:NQh})
above, one is left with the approximate system of coupled equations,
\begin{eqnarray}\label{eq:NGhbis}
\frac{d^2}{dY^{2}}N_g(Y)\!\!&\!\!=\!\!&\!\!
\gamma_0^2\left(N_s-a_1\frac{d}{dY}\right)N_g(Y),\\
\frac{d^2}{dY^{2}}N_q(Y)\!\!&\!\!=\!\!&\!\!\frac{C_F}{N_c}\gamma_0^2
\left(N_s-\tilde a_1\frac{d}{dY}\right)N_g(Y),\label{eq:NQhbis}
\end{eqnarray}
with the initial conditions at threshold $N_A^h(0)=1$ and $N_A^{'h}(0)=0$
and the hard constants
$$
a_1=\frac{1}{4N_c}\left[\frac{11}{3}N_c+
\frac43T_R\left(1-2\frac{C_F}{N_c}\right)\right],\qquad \tilde a_1=\frac34.
$$
The quantum corrections $\propto a_1,\;\tilde a_1$
in (\ref{eq:NGhbis},\ref{eq:NQhbis}) arise from the integration 
over the regular part of the splitting functions, they are 
${\cal O}(\sqrt\alpha_s)$ suppressed and {\em partially} account
for energy conservation as happens in the absence of the dense medium. Since
only the DLA terms are medium-enhanced in (\ref{eq:NGhbis},\ref{eq:NQhbis}), 
the hard constants are $N_s$-independent.

These equations can be solved
by applying the inverse Mellin transform:
\begin{equation}\label{eq:mellin}
N_g(Y)=\int_C\frac{d\omega}{2\pi i}e^{\omega Y'}N_g(\omega)
\end{equation}
to the self-contained gluonic equation (\ref{eq:NGhbis}), where
the contour $C$ lies to the right of all singularities of $N_G(\omega)$
in the complex plane. 
The running of the coupling constant $\alpha_s(Y)$, Eq. (\ref{eq:anodim}),
is taken into account through the identity
$$
\int_C\frac{d\omega}{2\pi i}Y'e^{\omega Y'}N_g(\omega)
=-\int_C\frac{d\omega}{2\pi i}e^{\omega Y'}\left(2\omega N_g(\omega)+
\omega^2\frac{d}{d\omega}N_g(\omega)\right).
$$
Consequently, one is left with
the following differential equation in Mellin space
\begin{equation}\label{eq:diffeq}
\frac1{N_g}\frac{d}{d\omega}N_g(\omega)=
-\frac{N_s}{\beta_0\omega^2}+\left(\frac{a_1}{\beta_0}-2\right)\frac1{\omega}.
\end{equation}
Solving (\ref{eq:diffeq}) and using (\ref{eq:mellin}), one obtains
\begin{equation}\label{eq:intrepres}
N_g(Y)\simeq\int_C\frac{d\omega}{2\pi i}\omega^{\frac{a_1}{\beta_0}-2}
\exp{\left(\omega Y'+\frac{N_s}{\beta_0\omega}\right)}.
\end{equation}
The exact solution of (\ref{eq:intrepres}) together with the initial
conditions leads to a linear combination
of two kinds of Bessel functions which resums the perturbative series at all powers
of $\sqrt\alpha_s$ \cite{Basics}. However, in this paper we are concerned with
the asymptotic solution of the equation
as $Y\gg1$ ($E\Theta\gg Q_0$), that is the high energy limit. 
Therefore, the Mellin transform (\ref{eq:intrepres}) can be estimated
by the steepest descent method. Indeed, the large parameter is $Y'$ and
the function in the exponent
presents a saddle point at $\omega_0=\sqrt{N_s/(\beta_0 Y')}$, such that 
the asymptotic solution reads
\begin{equation}\label{eq:nsmultg}
N_g^h(Y)\simeq {\cal K}\times 
{Y'}^{-\frac{\sigma_1}{\beta_0}}\exp{\sqrt{\frac{4N_s}{\beta_0}Y'}},
\end{equation}
where 
$$
\sigma_1=\frac{a_1}{2}-\frac{\beta_0}{4}.
$$
We also introduced, as stressed in the introduction, 
the LPHD normalization constant ${\cal K}$ \cite{LPHD}, which accounts
for hadronization effects outside the medium. 
The constant $\sigma_1$ is $N_s$-independent 
because it resums vacuum corrections. 
Therefore, the production of soft gluons in a dense medium becomes
$\exp\left[2(\sqrt{N_s}-1)\sqrt{Y'/\beta_0}\right]$ higher than the
standard production of soft gluons in the vacuum \cite{Basics}
and the factor $\sqrt{N_s}$ in
(\ref{eq:nsmultg}) underlines the
presence of the nuclear medium; this results has first 
been reported in \cite{perez}.
From (\ref{eq:nsmultg}) one obtains the first 
and second logarithmic derivatives of $N_g$: 
\begin{equation}\label{eq:logderiv}
\frac{1}{N_g}\frac{dN_g}{dY}\equiv\frac{1}{N_q}\frac{dN_q}{dY}=
\sqrt{N_s}\gamma_0-\sigma_1\gamma_0^2,\quad
\frac{1}{N_g}\frac{d^2N_g}{dY^2}\equiv\frac{1}{N_q}\frac{d^2N_q}{dY^2}=
N_s\gamma_0^2.
\end{equation}
The expression on the left hand side of 
(\ref{eq:logderiv}) is nothing but the MLLA rate
of multiparticle production with respect to the {\em evolution-time}
variable $Y\simeq\ln(\Theta)$ in the medium, which we
define as the medium-modified MLLA anomalous dimension:
\begin{equation}\label{eq:mllamedgam}
\gamma_{\text{med}}(Y)\equiv\frac1{N_g}\frac{dN_g}{dY}
=\sqrt{N_s}\gamma_0\left[1-\sigma_1\frac{\gamma_0}
{\sqrt{N_s}}+{\cal O}\left(\frac{\gamma_0^2}{N_s}\right)\right],
\end{equation}
where $N_s$ only affects, as expected, the leading double logarithmic term.
From (\ref{eq:nsmultg}) and (\ref{eq:mllamedgam}), one recovers the ansatz
\begin{equation}\label{eq:gammamed}
N_g(Y)\simeq\exp{\left(\int\gamma_{\text{med}}
(Y)dY\right)},
\end{equation}
which we further improve in the next paragraph by adding higher order
corrections. Finally, using (\ref{eq:nsmultg}) and 
(\ref{eq:NGhbis},\ref{eq:NQhbis}),
one obtains the solution for $N_Q^h$:
\begin{equation}\label{eq:nsmultq}
N_q(Y)=\frac{C_F}{N_c}\left[1+
\left(a_1-\tilde a_1\right)\frac{\gamma_0}{\sqrt{N_s}}\right]N_g(Y)
+{\cal O}\left(\frac{\gamma_0^2}{N_s}\right).
\end{equation}
Therefore, we can introduce the medium-modified MLLA
gluon to quark average multiplicity ratio
$r=N_g/N_q=N_g^h/N_q^h$ in the form
\begin{equation}\label{eq:mllaratio}
r=r_0\left[1-r_1\frac{\gamma_0}{\sqrt{N_s}}
+{\cal O}\left(\frac{\gamma_0^2}{N_s}\right)\right], \qquad
r_0=\frac{N_c}{C_F},\qquad r_1=a_1-\tilde a_1,
\end{equation}
where the suppression factor $1/\sqrt{N_s}$ 
restricts the production of hard collinear partons as $N_s>1$.
We notice that (\ref{eq:mllaratio}) is identical to the expression
with fixed coupling constant $\alpha_s(Y)$, where 
$Y=\ln\left(Q/Q_0\right)$ and $Q=E\Theta$ is
the virtuality of the jet produced in the nucleus-nucleus
reaction.
This factor is found to suppress the hard correction
${\cal O}(\gamma_0)$ and therefore, $r$ approaches
its asymptotic DLA limit $r_0=N_c/C_F$ when the coherent  
radiation of soft gluons is enhanced.
Setting $N_s=1$ in (\ref{eq:mllaratio}), one recovers the appropriate limits
in the vacuum \cite{Basics,MALAZA,KhozeOchs}. The constants entering in
(\ref{eq:nsmultg}) and (\ref{eq:mllaratio}) are
the same as those obtained in the vacuum and their values are displayed
in Table \ref{table:mllacoeff}. 
\begin{table}[htb]
\begin{center}
\begin{tabular}{cccccc}
\hline\hline
$n_f$ & $r_1$
& $\sigma_1$ \\ \hline
&&\\
3 & 0.185 & 0.280\\
4 & 0.191 & 0.297\\
5 & 0.198 & 0.314\\
&&\\
\hline\hline
\end{tabular}
\caption{Coefficients $r_1$ and $\sigma_1$.}
\label{table:mllacoeff}
\end{center}
\end{table}

\subsection{Medium-modified equations and solutions at Next-to-MLLA}
\label{subsec:nmllamult}
Previous MLLA results for the average multiplicities can be improved 
by further pushing the perturbative series in (\ref{eq:NGh},\ref{eq:NQh}).
We can include NNLO or NMLLA corrections of order
${\cal O}(\alpha_s)$, which are known to better account
for energy conservation in the vacuum \cite{DREMIN,DreminGary,PAM}.
For this purpose, we replace $N(Y+\ln x)$ 
($N(Y+\ln(1-x)$) by the Taylor expansion 
$N(Y)+\frac{d}{dY}N(Y)\ln x+\ldots$ 
($N(Y)+\frac{d}{dY}N(Y)\ln(1-x)+\ldots$) and, as
in section \ref{sub:eqmlla}, we integrate over the non-singular parts
of the splitting functions. We thus obtain the medium-modified NMLLA
approximate system of two-coupled evolution equations
\begin{eqnarray}\label{eq:NGhter}
\frac{d^2}{dY^{2}}N_g(Y)\!&\!=\!&\!
\gamma_0^2\left(N_s-a_1\left(\frac{d}{dY}-\beta\gamma_0^2\right)+
a_2(N_s)\frac{d^2}{dY^2}\right)N_g(Y),\\
\frac{d^2}{dY^{2}}N_q(Y)\!&\!=\!&\!\frac{C_F}{N_c}\gamma_0^2
\left(N_s-\tilde a_1\left(\frac{d}{dY}-\beta\gamma_0^2\right)+
\tilde a_2(N_s)\frac{d^2}{dY^2}\right)N_g(Y),\label{eq:NQhter}
\end{eqnarray}
with the new $N_s$-dependent constants
$$
a_2(N_s)=\frac{67}{36}-N_s\frac{\pi^2}{6}-\frac{13}{18}\frac{T_R}{N_c}
\frac{C_F}{N_c}+\frac23\frac{T_R}{N_c}\frac{C_F}{N_c}
\frac{(a_1-\tilde a_1)}{\sqrt{N_s}},\quad
\tilde a_2(N_s)=\frac{7}{8}+\frac{C_F}{N_c}
\left(\frac58-N_s\frac{\pi^2}{6}\right).
$$
The dependence of $a_2$ and $\tilde a_2$ on $N_s$ follows from the 
singular term in the integral
$N_s\int_0^1\frac{dx}{x}\ln(1-x)=-N_s\frac{\pi^2}{6}$ which
also enhances (see below) the induced soft gluon radiation
at the NNLO level. The term $\propto1/\sqrt{N_s}$ in $a_2(N_s)$
was obtained by replacing (\ref{eq:nsmultq}) in the single logarithmic
piece $\propto N_q$ in (\ref{eq:NGh}), while the term $\propto N_s$
in the equations (\ref{eq:NGhter},\ref{eq:NQhter}) enhances the
role of the leading DLA as in (\ref{eq:NGhbis},\ref{eq:NQhbis}).
As expected, one recovers the constants 
$a_2(N_s=1)=a_2$ and $\tilde a_2(N_s=1)
=\tilde a_2$ obtained in the vacuum \cite{PAM}
when $N_s=1$. The terms proportional
to these constants are known to better account for 
energy conservation in the partonic shower in the vacuum. 
Also note that the contributions $\propto a_1\beta_0$ and $\tilde a_1\beta_0$
cannot be neglected when performing predictions with running coupling
coupling constant.

The system (\ref{eq:NGhter},\ref{eq:NQhter}) can be solved by inserting the ansatz (\ref{eq:gammamed}) in both sides of (\ref{eq:NGhter}) with
\begin{equation}\label{eq:gamma}
\gamma_{\text{med}}(Y)
=\sqrt{N_s}\gamma_0\left[1-\sigma_1\frac{\gamma_0}
{\sqrt{N_s}}-\sigma_2(N_s)\frac{\gamma_0^2}{N_s}+{\cal O}\left(\frac{\gamma_0^3}{N_s^{3/2}}\right)\right],
\end{equation}
where $\sigma_2(N_s)$ is the unknown coefficient to be determined. 
The medium-modified NMLLA anomalous dimension
(\ref{eq:gamma}) has been inspired from the MLLA 
(\ref{eq:mllamedgam}) where, in both cases, we make appear
the rescaling of the coupling constant
$\alpha_s\to\alpha_s/N_s$ in the series. After equating 
terms $\propto\gamma_0^2$ in the left and right hand sides
of (\ref{eq:NGhter}) we obtain the following value
\begin{equation}\label{eq:sigma2}
\sigma_2(N_s)=-\frac12\left(\frac12a_1\beta_0+\frac14a_1^2+N_sa_2(N_s)+
\frac3{16}\beta_0^2\right),
\end{equation}
such that after integrating (\ref{eq:gamma}) according to (\ref{eq:gammamed}),
the medium-modified NMLLA average multiplicity takes the simple form
\begin{eqnarray}
\label{eq:nsmultgnmlla}
N_g^h(Y)\simeq{\cal K}\times{Y'}^{-\frac{\sigma_1}{\beta_0}}
\exp\left[{\sqrt{\frac{4N_s}{\beta_0}Y'}}+
\frac{2\sigma_2(N_s)}{\sqrt{N_s\beta^{3}_0Y'}}\right].
\end{eqnarray}
The term $\propto\sigma_2(N_s)$ in (\ref{eq:nsmultgnmlla}) 
provides the NMLLA correction to $N_G^h$.
As $N_s$ increases, $\sigma_2(N_s)$ follows the leading behavior 
$\sigma_2(N_s)\simeq N_s^2$ 
(and the second factor in the exponent
of (\ref{eq:nsmultgnmlla}) $\simeq N_s^{3/2}$) which
enhances the production of induced soft gluons in the medium at the NNLO level.
Setting $N_s=1$ in (\ref{eq:nsmultgnmlla}) one recovers
the value of this constant
in the vacuum 
$\sigma_2(N_s=1)\stackrel{n_f=3}{=}-0.378$ as given in \cite{DREMIN}.
Furthermore, setting $\beta_0=0$ in (\ref{eq:sigma2}), the appropriate
limit with fixed coupling constant $\alpha_s=const$ can be deduced.

We proceed to determine the medium-modified NMLLA gluon to quark average
multiplicity ratio by subtracting (\ref{eq:NQhter}) from (\ref{eq:NGhter}), one has
\begin{equation}\label{eq:substrac}
\frac{d^2}{dY^2}\left(N_g-\frac{N_c}{C_F}N_q\right)=
-(a_1-\tilde a_1)\frac{d}{dY}(\gamma_0^2N_g)
-(\tilde a_2(N_s)-a_2(N_s))\gamma_0^2
\frac{d^2N_g}{dY^2}.
\end{equation}
Applying the operators $\frac{d}{dY}$ and $\frac{d^2}{dY^2}$ to (\ref{eq:NQhter}) 
and dropping corrections contributing beyond NMLLA, we obtain
respectively
\begin{equation}\label{eq:step1}
N_s\frac{d}{dY}(\gamma_0^2N_g)=\frac{N_c}{C_F}\frac{d^3N_q}{dY^3}
+\tilde a_1\gamma_0^2\frac{d^2N_g}{dY^2},\quad
N_s\gamma_0^2\frac{d^2N_g}{dY^2}=\frac{N_c}{C_F}\frac{d^4N_q}{dY^4},
\end{equation}
where we keep track of the nuclear factor $N_s$ and the running of $\alpha_s(Y(x))$ 
in such a way that (\ref{eq:substrac}) can be rewritten 
in the form
\begin{equation*}
N_g\!=\!\frac{N_c}{C_F}N_q\!-\!
\frac{N_c}{C_F}\frac{(a_1\!-\!\tilde a_1)}{N_s}\frac{dN_q}{dY}
\!-\!\frac{N_c}{C_F}\!\left(\!\frac{\tilde a_1(a_1\!-\!\tilde a_1)}{N_s^2}\!+\!
\frac{(\tilde a_2(N_s)\!-\!a_2(N_s))}{N_s}\!\right)\!\frac{d^2N_q}
{dY^2},
\end{equation*}
after reassembling
terms $\propto\frac{dN_q}{dY}$ and 
$\propto\frac{d^2N_q}{dY^2}$.
Using (\ref{eq:logderiv}) together with the initial 
conditions at threshold yields for $r=N_g/N_q$ with NNLO accuracy
\begin{equation}\label{eq:nmllaratio}
r=r_0\left[1-r_{1}\frac{\gamma_0}{\sqrt{N_s}}-r_2(N_s)
\frac{\gamma_0^2}{N_s}+{\cal O}\left(\frac{\gamma_0^3}{N_s^{3/2}}\right)\right],
\end{equation}
where the coefficient $r_2$ is explicitly dependent on $N_s$
through the formula
\begin{eqnarray}
r_2(N_s)=\left(\tilde a_1-\sigma_1\right)r_1
+(\tilde a_2(N_s)-a_2(N_s))N_s.
\end{eqnarray}
The dependence of $r_2(N_s)$ on $\beta_0$ in (\ref{eq:nmllaratio}) underlines
the account of the running coupling constant, setting $\beta_0=0$
the appropriate fixed coupling solution can be deduced.
For $N_s=1$, the appropriate limits in the vacuum are 
recovered \cite{DREMIN}, for example
$$
r_2(N_s=1)=\left(\tilde a_1-\sigma_1\right)r_1
+(\tilde a_2(1)-a_2(1))\stackrel{n_f=3}{=}0.426.
$$

We give the NMLLA coefficients $\sigma_2(N_s)$ and $r_2(N_s)$ defined in 
(\ref{eq:nsmultgnmlla}) and (\ref{eq:nmllaratio}) as a function of $N_s$ 
in Table \ref{table:coeff}.

\begin{table}[htb]
\begin{center}
\begin{tabular}{cccccc}
\hline
\hline
$n_f$ & $r_2(N_s)$
& $\sigma_2(N_s)$ \\ \hline
&&\\
3 & $0.087-0.027\sqrt{N_s}-0.548N_s+0.914N_s^2$ & $-0.337-0.014\sqrt{N_s}-0.850N_s + 0.822N_s^2$\\
&&\\
4 & $0.087-0.038\sqrt{N_s}-0.494N_s+0.914N_s^2$ & $-0.319-0.019\sqrt{N_s}-0.823N_s + 0.822N_s^2$\\
&&\\
5 & $0.087-0.049\sqrt{N_s}-0.441N_s+0.914N_s^2$ & $-0.302-0.024\sqrt{N_s}-0.797N_s + 0.822N_s^2$\\
&&\\
\hline\hline
\end{tabular}
\caption{Coefficients $r_2(N_s)$ and $\sigma_2(N_s)$.}
\label{table:coeff}
\end{center}
\end{table}
As $N_s$ increases, the ${\cal O}(\gamma_0/\sqrt{N_s})$ correction decreases,
while the one ${\cal O}(\gamma_0^2/N_s)$ becomes sizable and decreases 
like $\simeq-N_s$. That is why it might be wondered whether the 
convergence of the perturbative series could be 
reached at a certain level of accuracy. Since the series widely 
oscillate at low energy scales, large terms 
$\propto\pi^2$ in $a_2(N_s)$ and $\tilde a_2(N_s)$ might spoil
or drastically affect the trends obtained at MLLA. 
This kind of behavior
has first been noticed in the Koba-Nielsen-Olsen (KNO) 
problem \cite{DokKNO} in the vacuum.

\subsection{NLO and NNLO results on $\boldsymbol{N_g^h}$ and $\boldsymbol{r}$}

\label{subsec:figs1}

\begin{figure}[t]
\begin{center}
\epsfig{file=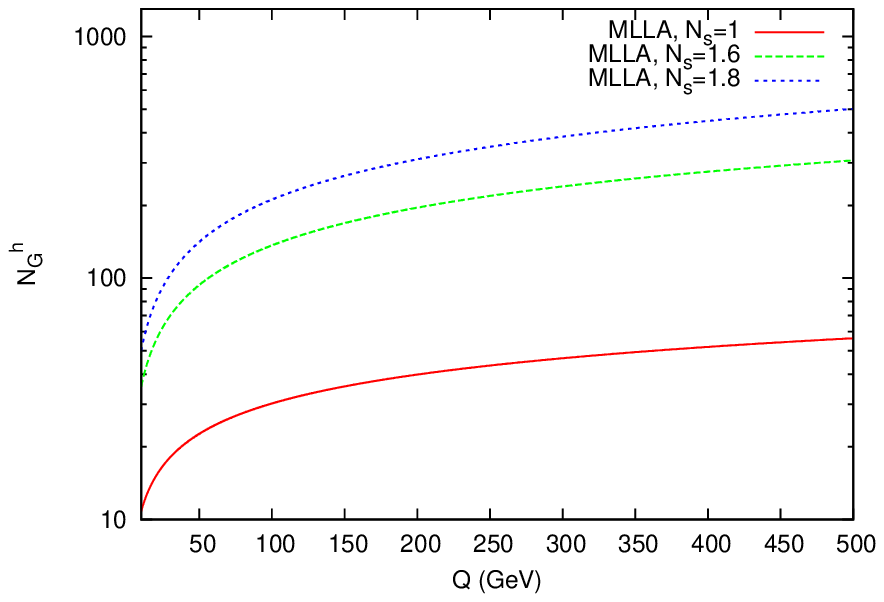,width=7.5truecm}
\qquad
\epsfig{file=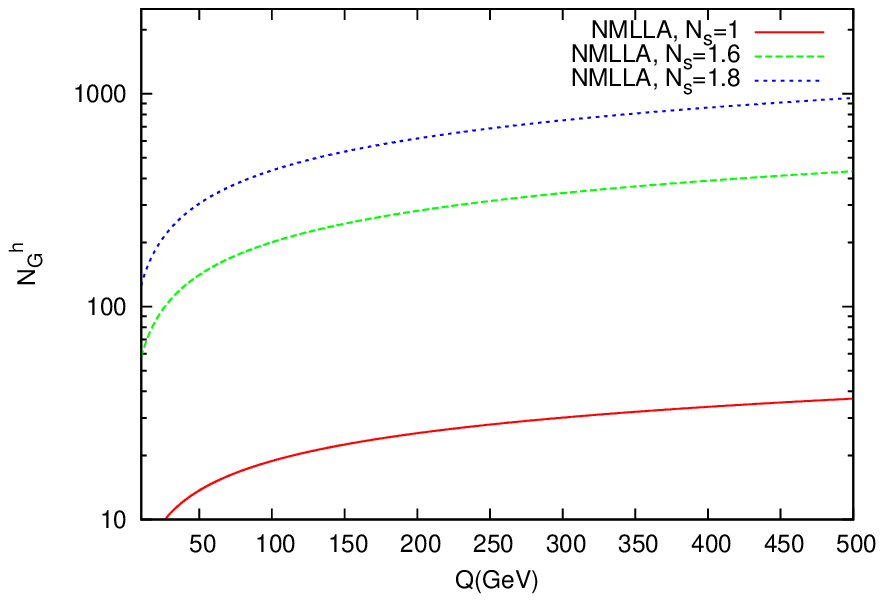,width=7.5truecm}
\caption{MLLA (\ref{eq:nsmultg}) 
and NMLLA (\ref{eq:nsmultgnmlla}) average multiplicity as a function
of $Q=E\Theta$ in the vacuum ($N_s=1$) and
in the medium ($N_s=1.6$ and $N_s=1.8$).
\label{fig:avemult}
}
\end{center}
\end{figure}

Setting $n_f=3$, we display in Fig.\,\ref{fig:avemult} 
our results for the medium-modified MLLA
(\ref{eq:nsmultg}) and NMLLA (\ref{eq:nsmultgnmlla}) average 
multiplicity. 
We plot $N_g^h$ in the range $10\leq Q(\text{GeV})\leq500$, where $Q=E\Theta$ is
the total virtuality of the jet related to $Y$ in 
(\ref{eq:anodim}). We compare our results in the medium 
for $N_s=1.6$ and $N_s=1.8$ (see \cite{Urs}) 
with predictions in the vacuum ($N_s=1$), we set
$Q_0=\Lambda_{QCD}=0.23$ GeV in the limiting spectrum
approximation \cite{DREMIN}, and ${\cal K}=0.2$ is taken
from \cite{DREMIN}. The values $N_s=1.6$ and $N_s=1.8$
in the medium may be realistic for RHIC and LHC phenomenology
\cite{Urs,sapetabis}; the jet energy subrange $10\leq Q(\text{GeV})\leq50$ displayed in
Fig.\,\ref{fig:avemult} has been recently considered by the STAR
collaboration, which
reported the first measurements of charged hadrons and
particle-identified
fragmentation functions from p+p collisions \cite{heinz} at
$\sqrt{s_{\text{NN}}}=200$ GeV. Finally, the jet energy range
in the same figure,
in particular for those values at $Q\geq50$ GeV, will
be reached at the LHC, i.e $Q=100$ GeV is an
accessible value in this experiment (see \cite{Urs}
and references therein).

Notice that at NMLLA, the increase of
$N_g^h$ with $N_s$ is more substantial
than at MLLA. The former is driven by the leading contribution
to $\sigma_2(N_s)$: it
increases like $\sigma_2(N_s)\simeq N_s^2$ 
(see Table \ref{table:coeff}) in the sub-leading 
piece of (\ref{eq:nsmultgnmlla}). In both resummations schemes we find, 
as expected from our calculations, that the production of 
soft hadrons increases as $N_s>1$, which implies that the available phase space
becomes restricted for the production of harder collinear hadrons.

\begin{figure}[t]
\begin{center}
\epsfig{file=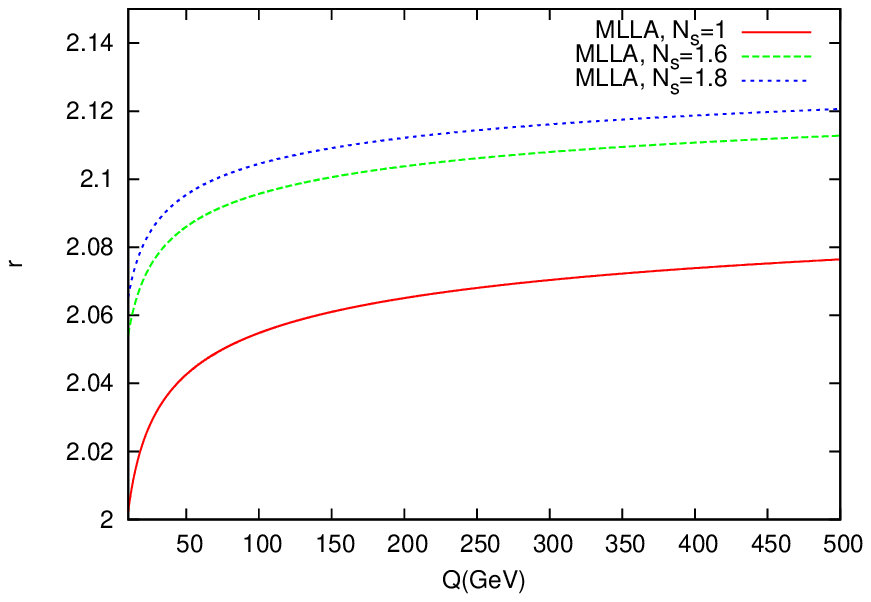,width=7.5truecm}
\qquad
\epsfig{file=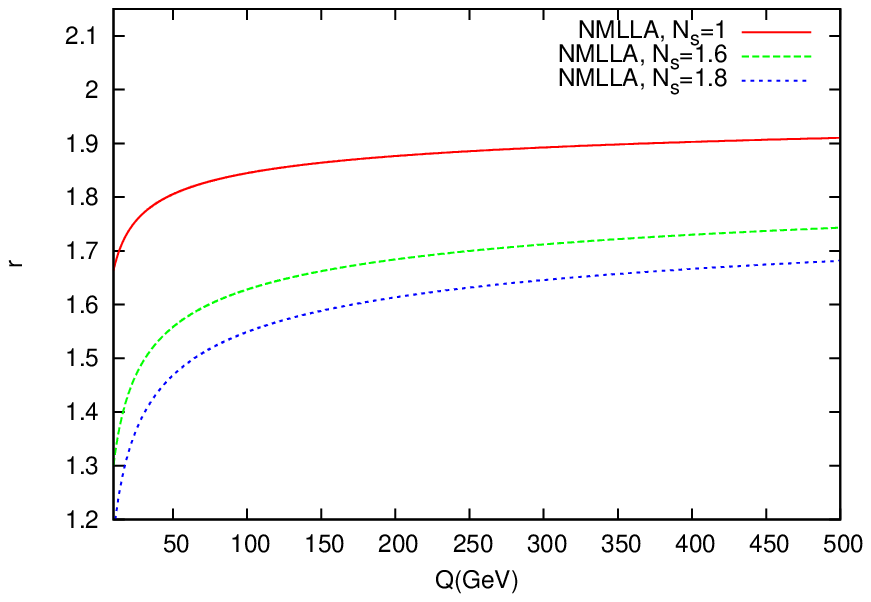,width=7.5truecm}
\caption{MLLA (\ref{eq:mllaratio}) 
and NMLLA (\ref{eq:nmllaratio}) gluon to quark average multiplicity ratio as a function
of $Q=E\Theta$ in the vacuum ($N_s=1$) and
in the medium ($N_s=1.6$ and $N_s=1.8$).
\label{fig:ratio}
}
\end{center}
\end{figure}

In Fig.\,\ref{fig:ratio}, we display the medium-modified MLLA ratio $r=N_g/N_q$ 
(\ref{eq:mllaratio}) and the medium-modified
NMLLA ratio as a function of $Q=E\Theta$. 
As expected from (\ref{eq:mllaratio}), as $N_s$
increases, the ${\cal O}(\gamma_0)$ correction is suppressed by 
$1/\sqrt{N_s}$, the ratio approaches the DLA asymptotic regime 
$r_0=N_c/C_F=9/4$.
At NMLLA, the previous trend goes in the opposite direction:
one has indeed $-r_2(N_s)\simeq-N_s^2$ (see Table \ref{table:coeff}) 
in the ${\cal O}(\gamma_0^2/N_s)$ piece of
(\ref{eq:nmllaratio}), which is negative and sizable
as $N_s$ increases and, therefore, spoils the behavior obtained
at MLLA signaling difficulties with 
perturbative theory in the medium. 
Indeed, sizable oscillations have been noticed in the 
perturbative series \cite{dln}
and it turns out that they are wider in the medium than in the vacuum. For
example, the NMLLA correction to $r$ is $\sim10\%$ for
$N_s=1$, and for $N_s=1.8$ it is $\sim40\%$. 
It should be noticed that
every logarithmic derivative of $N_A^h$ provides a half power of $N_s$
to successive terms in the series in the form
$\frac{d^nN}{dY^n}\approx(N_s\alpha_s)^{n/2}$ 
($n=1\to$ MLLA, $n=2\to$ NMLLA\ldots), such that the 
perturbative approach should fail 
as higher order terms are
incorporated.
Former statements suggest that by incorporating NMLLA 
and next-to-NMLLA (NNMLLA, see paragraph \ref{subsec:higher})  
corrections on an equal footing, 
the MLLA behavior can be recovered. We conclude
from this analysis that 
MLLA provides a more realistic
physical picture of the softening of jets 
than NMLLA. Therefore, either
the incorporation of NNMLLA terms or the exact numerical 
solution of the evolution equations \cite{LupiaOchs,sapeta}, which
exactly accounts for the running of $\alpha_s$ and the energy balance, is
required.
In \cite{termo}, a numerical solution
of the equations was provided with fixed coupling constant $\alpha_s$, and the results
are shown to follow our MLLA expectations as $N_s$ increases. 

Finally, in both MLLA and NMLLA, the gluon jets are still more active
than the quark jets in producing secondary particles  and the shape of the curves
are roughly identical; however, these 
characteristics prove not to
be very sensitive to $N_s$.

\section{Medium-modified evolution for the second multiplicity correlator}

\label{section:2corr}

The second multiplicity correlator 
was first considered in \cite{MALAZA} at MLLA and later
in \cite{dln} at NMLLA. 
It is defined in the form $N_A^{(2)}=\langle {N_A}({N_A}-1)\rangle$
in gluon ($A=g$) and quark ($A=q$) jets. 
The normalized second multiplicity
correlator defines the width of the multiplicity distribution and is related to its 
dispersion $D_A^2=\langle {N_A}({N_A}-1)\rangle-N_A^2$ 
by the formula
\begin{equation}
D_A^2=(A_2-1)N_A^2+N_A.
\end{equation}
The second multiplicity correlators normalized to their own squared
average multiplicity are
\begin{equation}
G_2=\frac{\langle {N_g}({N_g}-1)\rangle}{N_g^2},\quad 
Q_2=\frac{\langle {N_q}({N_q}-1)\rangle}{N_q^2},
\end{equation}
inside a gluon and a quark jet respectively.
These observables are obtained by integrating the double differential inclusive 
cross section over the energy fractions $x_1=e_1/E$ and $x_2=e_2/E$
$$
\langle {N_A}({N_A}-1)\rangle=\iint dx_1dx_2\left(\frac1{\sigma}
\frac{d^2\sigma}{dx_1dx_2}\right)_A.
$$
The correlators are ${\cal K}$-independent and provide a pure test of
multiparticle dynamics. However, important disagreements with $e^+e^-$
data \cite{experN2} indicates that
non-trivial hadronization effects may play a significant role. The treatment of this
observable with full account of perturbative and non-perturbative ingredients
is not available yet. Therefore, we
study the variation of this observable as $N_s>1$ with respect to the
limit in the vacuum $N_s=1$ and do not compare our results 
with $e^+e^-$ data \cite{experN2}.
The medium-modified system of two-coupled evolution equations for this observable
follows from the MLLA master equation for the azimuthally averaged generating
functional and can be written in the convenient form

\vbox{
\begin{eqnarray}
\frac{d}{dY}(N_g^{(2)}-N_g^2)\!\!&\!\!=\!\!&\!\!\int_0^1dx\gamma_0^2\Phi_g^g\left[N_g^{(2)}
(Y+\ln x)\!+\!\Big(N_g^{(2)}(Y+\ln(1-x))-N_g^{(2)}(Y)\Big)\right.\cr
&&\left.\hskip 0.5cm+\Big(N_g(Y+\ln x)-N_g(Y)\Big)
\Big(N_g(Y+\ln(1-x))-N_g(Y)\Big)\right]\cr
&&\hskip -1cm+n_f\!\!\int_0^1dx\gamma_0^2\Phi_g^q
\left[2\Big(N_q^{(2)}(Y+\ln x)\!-\!N_q^2(Y+\ln x)\Big)
\!-\!\Big(N_g^{(2)}(Y)\!-\!
N_g^2(Y)\Big)\right.\cr
\!\!&\!\!+\!\!&\!\!\left.\Big(2N_q(Y+\ln x)-N_g(Y)\Big)
\Big(2N_q(Y+\ln(1-x))-N_g(Y)\Big)\right]\label{eq:corrg},\\\cr
\frac{d}{dY}(N_q^{(2)}-N_q^2)\!\!&\!\!=\!\!&\!\!\int_0^1dx\gamma_0^2
\Phi_q^g
\left[N_g^{(2)}(Y+\ln x)\!+\!\left(N_q^{(2)}(Y+\ln(1-x))-N_q^{(2)}(Y)\right)\right.\cr
&&\left.\hskip 0.3cm+2\Big(N_g(Y+\ln x)- N_q(Y)\Big)
\Big(N_q(Y+\ln(1-x))- N_q(Y)\Big)\right]\label{eq:corrq},
\end{eqnarray}
}
which proves to be more suitable for obtaining analytical solutions
in the following. We use a new method to compute solutions at MLLA and NMLLA by replacing
$N_A^{(2)}=A_2N_A^2$ on both sides of the expanded equations at 
$x\sim1-x\sim1$. This
observable is less inclusive than the average multiplicity, it can
indeed be derived
from the two-particle four-momentum correlation \cite{RPR} in the shower. The
medium-modified formul{\ae}  
(\ref{eq:gammamed}), (\ref{eq:mllaratio}) and (\ref{eq:nmllaratio}) will be used
in this analysis.
\subsection{MLLA approximation}

\label{subsec:2multmlla}

For $Y\gg\ln x\sim\ln(1-x)$ in the system above 
(\ref{eq:corrg},\ref{eq:corrq}), 
$N(Y+\ln x)$ ($N(Y+\ln(1-x))$) and 
$N^{(2)}(Y+\ln x)$ ($N^{(2)}(Y+\ln(1-x))$) can be
replaced by $N(Y)$ and $N^{(2)}(Y)$ respectively
 in the hard partonic splitting region
$x\sim1-x\sim1$, while  the dependence on $\ln x$ is kept on the singular
one ($x\to0$). 
The medium-modified MLLA approximate system of two-coupled
evolution equations for the second multiplicity correlator
reads
\begin{eqnarray}\label{eq:N2G}
\frac{d^2}{dY^2}\left(N^{(2)}_g-N_g^2\right)
\!\!&\!\!=\!\!&\!\!\gamma_0^2\left(N_s-a_1\frac{d}{dY}\right)N^{(2)}_g
+(a_1-b_1)\gamma_0^2\frac{d}{dY} N_g^2,\\
\frac{d^2}{dY^2}\left(N^{(2)}_q-N_q^2\right)\!\!&\!\!=\!\!&\!\!
\frac{C_F}{N_c}\gamma_0^2\left(N_s-
\tilde a_1\frac{d}{dY}\right)N^{(2)}_g,
\label{eq:N2Q}
\end{eqnarray}
where the new hard constant is:
$$
b_1=\frac1{4N_c}\left[
\frac{11}{3}N_c-4\frac{T_R}{N_c}\left(1-2\frac{C_F}{N_c}\right)^2\right].
$$ 
The constant $N_s$ only affects the leading double logarithmic term of the
equations. The terms proportional to $a_1$, $(a_1-b_1)$ and $\tilde a_1$ are
hard vacuum corrections which {\em partially} account for energy conservation, indeed
$\gamma_0^2\frac{dN}{dY}\approx\sqrt{N_s}\gamma_0^3$ and the relative correction
to DLA is ${\cal O}(\sqrt{\alpha_s/N_s})$. 
As before, the hard constants are, as expected, $N_s$-independent.
Moreover, these sub-leading contributions 
have the same form as those describing 
the two-particle correlation 
\cite{RPR}. 

\subsubsection{Medium-modified $\boldsymbol{G_2}$ at MLLA
and expansion in $\boldsymbol{{\cal O}(\gamma_0/\sqrt{N_s})}$}

Setting $N^{(2)}_g=G_2N_g^2$ in (\ref{eq:N2G}), the system can be solved 
iteratively by considering terms up to ${\cal O}(\alpha_s^{3/2})$ in the
left and right hand sides of (\ref{eq:N2G}). 
Thus, the l.h.s. of (\ref{eq:N2G}) writes in the form
\begin{equation}\label{eq:lhs}
\frac{d^2}{dY^2}\left(N^{(2)}_g- N_g^2\right)
=\frac{d^2G_2}{dY^2} N_g^2+2\frac{dG_2}{dY}
\frac{d}{dY} N_g^2+(G_2-1)\frac{d^2}{dY^2}N_g^2.
\end{equation}
Hereafter, in all sub-leading terms, we can replace $G_2$ by a constant
$G_2=G_2^{\text{DLA}}=const$, while the terms involving $N_g$ should be computed 
by using (\ref{eq:nsmultg}), thus
\begin{eqnarray}\label{eq:lhsg}
\frac{d^2}{dY^2}\left(N^{(2)}_g- N_g^2\right)
=\gamma_0^2N_s(G_2-1)\left[4-(4a_1-\beta_0)\frac{\gamma_0}{\sqrt{N_s}}\right]
N_g^2
\end{eqnarray}
while the r.h.s. reads
\begin{eqnarray}\label{eq:rhsg}
\left[\left(N_s\gamma_0^2-a_1\gamma_0^2\frac{d}{dY}\right)G_2
+(a_1-b_1)\gamma_0^2\frac{d}{dY}\right] N_g^2&=&\\
&&\hskip -6cm\gamma_0^2N_s(G_2-1)N_g^2+N_s\gamma_0^2\left[1-\left(\frac23a_1+2b_1\right)
\frac{\gamma_0}{\sqrt{N_s}}\right] N_g^2.\notag
\end{eqnarray}
Equating (\ref{eq:lhsg}) and (\ref{eq:rhsg}) the exact MLLA solution
of (\ref{eq:N2G}) reads
\begin{equation}\label{eq:G2MLLA}
G_2-1=
\frac{1-\delta_1\displaystyle{\frac{\gamma_0}{\sqrt{N_s}}}}{3-\delta_2\displaystyle{\frac{\gamma_0}{\sqrt{N_s}}}},
\end{equation}
where 
$$
\delta_1=\left(\frac23a_1+2b_1\right),\quad
\delta_2=(4a_1-\beta_0).
$$
Setting $\gamma_0/\sqrt{N_s}\to0$ in (\ref{eq:G2MLLA}) one recovers the DLA
$N_s$-independent value $G^{\text{DLA}}_2=4/3$. 
Then, expanding (\ref{eq:G2MLLA}) in the form $1+\gamma_0/\sqrt{N_s}$ , 
one recovers the result from \cite{MALAZA} for $N_s=1$
\begin{equation}\label{eq:expG2}
G_2-1\approx\frac13-c_1\frac{\gamma_0}{\sqrt{N_s}}+
{\cal O}\left(\frac{\gamma_0^2}{N_s}\right),
\end{equation}
where the linear combination of color factors reads
\begin{equation}\label{eq:c1}
c_1=-\frac29a_1+\frac19\beta_0+\frac23b_1
=\frac{1}{4N_c}\!\left(\!\frac{55}{9}-4\frac{T_R}{N_c}
+\frac{112}{9}\frac{T_R}{N_c}\frac{C_F}{N_c}
-\frac{32}{3}\frac{T_R}{N_c}\frac{C_F^2}{N_c^2}\right).
\end{equation}

\subsubsection{Medium-modified $\boldsymbol{Q_2}$ at MLLA and 
expansion in $\boldsymbol{{\cal O}(\gamma_0/\sqrt{N_s})}$}

Inserting (\ref{eq:G2MLLA}), $N_q^{(2)}=Q_2 N_q^2$ and using the
MLLA expression for the ratio (\ref{eq:nsmultq}) in (\ref{eq:N2Q}), 
it is straightforward to obtain
\begin{equation}\label{eq:Q2MLLA}
\frac{Q_2-1}{G_2-1}=\frac{N_c}{C_F}\left[1+\frac{3}{2}(b_1-a_1)
\frac{\gamma_0}{\sqrt{N_s}}+{\cal O}\left(\frac{\gamma_0^2}{N_s}\right)\right].
\end{equation}
Expanding (\ref{eq:Q2MLLA}) in the 
form $1+\gamma_0/\sqrt{N_s}$ and setting $N_s=1$, one recovers the
result from \cite{MALAZA} in the vacuum. Indeed,
\begin{eqnarray}
Q_2-1\approx\frac{N_c}{C_F}\left(\frac1{3}-\tilde c_1\frac{\gamma_0}{\sqrt{N_s}}\right)
+{\cal O}\left(\frac{\gamma_0^2}{N_s}\right),
\label{eq:expQ2}
\end{eqnarray}
where, in agreement with \cite{MALAZA},
we obtain the combination of color factors
\begin{eqnarray}
\tilde c_1=\frac5{18}a_1+\frac16b_1
+\frac19\beta_0=\frac{1}{4N_c}\!
\left(\!\frac{55}{9}+\frac49\frac{T_R}{N_c}\frac{C_F}{N_c}
-\frac83\frac{T_R}{N_c}\frac{C_F^2}{N_c^2}\right).
\end{eqnarray}
In Table \ref{table:c-one} we display $\tilde c_1$ together with  
$c_1$ (\ref{eq:c1}) for $n_f=3,4,5$. 
\begin{table}[htb]
\begin{center}
\begin{tabular}{cccccc}
\hline\hline
$n_f$ & $c_1$
& $\tilde c_1$ \\ \hline
&&\\
3 & 0.485 & 0.495\\
4 & 0.477 & 0.491\\
5 & 0.469 & 0.486\\
&&\\
\hline\hline
\end{tabular}
\caption{Coefficients $c_1$ and $\tilde c_1$.}
\label{table:c-one}
\end{center}
\end{table}
As for the medium-modified MLLA expression 
$r=N_g/N_q$ (\ref{eq:mllaratio}), 
hard corrections to the MLLA second multiplicity 
correlators $G_2$ and $Q_2$ are suppressed by the factor 
$1/\sqrt{N_s}$, while the leading double logarithmic terms ($\gamma_0/\sqrt{N_s}\to0$) remain
unchanged and equal the vacuum result
\begin{equation}
A_2=1+\frac{N_c}{3C_A}, \quad A=g\;(C_g=N_c), \quad A=q\;(C_q=C_F).
\end{equation}
Thus, our MLLA predictions for the medium-modified second multiplicity
correlators follow the characteristics of the jet quenching.

\subsection{Next-to-MLLA evolution equations for the multiplicity correlator}

To obtain the equations we proceed like in paragraph \ref{subsec:nmllamult} and use
results from subsection \ref{subsec:2multmlla}. Indeed, by further pushing
the perturbative series, one can improve the account of the energy balance. 
We replace, in the hard splitting region $Y\gg\ln x\sim\ln(1-x)$, 
$N(Y+\ln x)$ ($N(Y+\ln(1-x))$) and 
$N^{(2)}(Y+\ln x)$ ($N^{(2)}(Y+\ln(1-x))$) by 
$N(Y)+\frac{d}{dY}N(Y)\ln x\ldots$ 
($N(Y)+\frac{d}{dY}N(Y)\ln(1-x)\ldots$) and 
$N^{(2)}(Y)+\frac{d}{dY}N^{(2)}(Y)\ln x\ldots$ 
($N^{(2)}(Y)+\frac{d}{dY}N^{(2)}(Y)\ln(1-x)\ldots$) 
respectively in the system (\ref{eq:corrg},\ref{eq:corrq}), 
while  the dependence on $\ln x$ is kept on the singular piece $N_s/x$. 
After integrating the regular terms over $x$, 
the medium-modified NMLLA approximate system of two-coupled
evolution equations for the
gluon and quark multiplicity correlator reads
\begin{eqnarray}\label{eq:NN2G}
\frac{d^2}{dY^2}\left(N^{(2)}_g-N_g^2\right)
\!\!&\!\!=\!\!&\!\!\gamma_0^2\left(N_s-a_1\left(\frac{d}{dY}-
\beta_0\gamma_0^2\right)
+a'_2(N_s)\frac{d^2}{dY^2}\right)\!N^{(2)}_g\\
\!\!&\!\!+\!\!&\!\!\gamma_0^2\left((a_1-b_1)\left(\frac{d}{dY}-
\beta_0\gamma_0^2\right)+b_2(N_s)\frac{d^2}{dY^2}\right)\!N_g^2\notag\\
\!\!&\!\!+\!\!&\!\!\gamma_0^3\, b_3(N_s)\frac{d}{dY}\left(N_g^{(2)}-N_g^2\right),\notag\\
\frac{d^2}{dY^2}\left(N^{(2)}_q-N_q^2\right)\!\!&\!\!=\!\!&\!\!
\frac{C_F}{N_c}\gamma_0^2\left(N_s-
\tilde a_1\left(\frac{d}{dY}-\beta_0\gamma_0^2\right)+
\tilde a_2(N_s)\frac{d^2}{dY^2}\right)N^{(2)}_g,
\label{eq:NN2Q}
\end{eqnarray}
where the term $\propto b_3$ follows from the MLLA result (\ref{eq:Q2MLLA}),
$$
N_q^{(2)}-N_q^2=\frac{C_F}{N_c}\left[1+
\left(\frac12a_1-2\tilde a_1+
\frac32b_1\right)\frac{\gamma_0}{\sqrt{N_s}}
\right]\left(N_g^{(2)}-N_g^2\right).
$$
The constants are the following:
\begin{eqnarray}
a'_2(N_s)\!\!&\!\!=\!\!&\!\! a_2(N_s)-\frac{2T_R}{3N_c}\frac{C_F}{N_c}
\frac{r_1}{\sqrt{N_s}},\\
b_2(N_s)\!\!&\!\!=\!\!&\!\! \frac{T_R}{3N_c}\frac{C_F}{N_c}
\left[\frac{13}{3}\left(1-\frac{C_F}{N_c}\right)-2\left(1-2\frac{C_F}{N_c}\right)
\frac{r_1}{\sqrt{N_s}}\right],\\
b_3(N_s)\!\!&\!\!=\!\!&\!\!\frac{T_R}{3N_c}\frac{C_F}{N_c}
\left(\frac{r_1}{\sqrt{N_s}}+3\frac{(b_1-\tilde a_1)}{\sqrt{N_s}}\right).
\end{eqnarray}
The terms $\propto a_1\beta_0,\;a_2'(N_s),\;(a_1-b_1)\beta_0,\;b_2(N_s),\;b_3(N_s)$
in (\ref{eq:NN2G}) and the ones $\propto \tilde a_1\beta_0,\;\tilde a_2(N_s)$ in
(\ref{eq:NN2Q}) are ${\cal O}(\gamma_0^2)$ corrections which 
better account for energy conservation. We remind
that $\frac{d^nN}{dY^n}\simeq{\cal O}((N_s\alpha_s)^{n/2})$ and 
that terms $\propto\beta_0$
arise from the running of the coupling constant $\alpha_s(Y)$. 
Moreover, these constants
take $N_s$-dependence for 
the reasons explained in section
\ref{subsec:nmllamult}.

\subsubsection{Medium-modified $\boldsymbol{G_2}$ at NMLLA and 
expansion in $\boldsymbol{{\cal O}(\gamma_0/\sqrt{N_s})}$}

Setting $N_g^{(2)}=G_2N_g^2$ in (\ref{eq:NN2G}), the equation
can be solved iteratively by making use of (\ref{eq:gamma}),
the MLLA formula (\ref{eq:expG2}) for $G_2$
and the leading DLA limit $G^{\text{DLA}}_2=4/3$; moreover, we expand
the series up to terms ${\cal O}(\alpha_s^2)$.
The l.h.s. of (\ref{eq:NN2G}) can therefore
be written in the form,
\begin{eqnarray}\label{eq:lhsnmlla}
l.h.s.\!\!&\!\!=\!\!&\!\! \gamma_0^2(G_2-1)
\left[4N_s-(8\sigma_1+\beta_0)\sqrt{N_s}\gamma_0
+2\left(\sigma_1(2\sigma_1+\beta_0)-4\sigma_2(N_s)\right)
\gamma_0^2\right]N_g^2\\
\!\!&\!\!+\!\!&\!\!2\beta_0c_1\gamma_0^4N_g^2.\notag
\end{eqnarray}
The r.h.s. reads
\begin{eqnarray}\label{eq:rhsnmlla}
r.h.s.\!\!&\!\!=\!\!&\!\!\gamma_0^2\left[N_s(G_2-1)+N_s
-2\left(\frac{1}{3}a_1+b_1\right)\sqrt{N_s}\gamma_0+
\left(\frac{1}{3}a_1+b_1\right)(2\sigma_1+\beta_0)\gamma_0^2\right.\\
\!\!&\!\!+\!\!&\!\!\left.2\left(\frac{1}{3}b_3(N_s)+a_1\frac{c_1}{\sqrt{N_s}}\right)
\sqrt{N_s}\gamma_0^2+4\left(\frac43a'_2(N_s)+b_2(N_s)\right)N_s\gamma_0^2\right]N_g^2.
\notag
\end{eqnarray}
Equating (\ref{eq:lhsnmlla}) and (\ref{eq:rhsnmlla}) we find the new exact NMLLA
solution of (\ref{eq:NN2G}),
\begin{eqnarray}\label{eq:solg2}
G_2-1=\frac{1-\delta_1\displaystyle{\frac{\gamma_0}{\sqrt{N_s}}}+\delta_3(N_s)
\displaystyle{\frac{\gamma_0^2}{N_s}}}
{3-\delta_2\displaystyle{\frac{\gamma_0}{\sqrt{N_s}}}+
\delta_4(N_s)\displaystyle{\frac{\gamma_0^2}{N_s}}},
\end{eqnarray}
where the following combinations of color factors have been written
in the form
\begin{eqnarray}
\delta_3(N_s)\!\!&\!\!=\!\!&\!\!\left(\frac{1}{3}a_1+b_1\right)(2\sigma_1+\beta_0)+2\left(\frac{1}{3}b_3(N_s)
+\frac{c_1}{\sqrt{N_s}}\left(a_1-\beta_0\right)\right)\sqrt{N_s}\\
\!\!&\!\!+\!\!&\!\!4\left(\frac43a'_2(N_s)+b_2(N_s)\right)N_s,\\
\delta_4(N_s)\!\!&\!\!=\!\!&\!\!2\left(\sigma_1(2\sigma_1+\beta_0)
-4\sigma_2(N_s)\right).
\end{eqnarray}
When the MLLA coefficients $\delta_1$, $\delta_2$ and NMLLA $\delta_3(N_s)$ and 
$\delta_4(N_s)$ are evaluated in the vacuum ($N_s=1$) for $n_f=3$,
we find respectively $\delta_1=2.453$, $\delta_2=2.991$, $\delta_3(1)=2.818$ 
and $\delta_4(1)=3.766$. In particular, $\delta_1\sim\delta_3(1)$
while the NMLLA $\delta_4(1)$ becomes bigger than the MLLA $\delta_2$. It
was shown in the KNO problem that
MLLA corrections increase like $\sim k\sqrt\alpha_s$ ($k=2$) 
while NMLLA like $\sim k^2\alpha_s$ ($k^2=4$) as the rank of the correlator, which
coincides with the number of particles triggered in the shower,
increases \cite{DokKNO}. It may be the reason why
sizable NMLLA coefficients 
are found in this picture.
Moreover, as the rank $k$ 
of the correlator increases, ${\cal O}(\sqrt\alpha_s)$ corrections
become of the same order of magnitude than
the leading DLA and perturbation theory fails. 
Therefore and in general, MLLA and NMLLA corrections for the
less inclusive multiplicity correlator of any 
rank $k$ are more sizable than those 
of the more inclusive average multiplicity. 
That is the reason for, the exact numerical solution of the 
evolution equations \cite{LupiaOchs,sapeta} becomes interesting. 

Expanding (\ref{eq:solg2}) in $\gamma_0/\sqrt{N_s}$ in the form
$1+\gamma_0/\sqrt{N_s}+\gamma_0^2/N_s$, we obtain
\begin{equation}\label{eq:expandG2}
G_2-1=\frac13-c_1\frac{\gamma_0}{\sqrt{N_s}}+
c_2(N_s)\frac{\gamma_0^2}{N_s}+{\cal O}\left(\frac{\gamma_0^3}{N_s^{3/2}}\right)
\end{equation}
where
\begin{equation}\label{eq:c2}
c_2(N_s)=\frac1{27}\left(\delta_2(N_s)^2-3\delta_4(N_s)
-3\delta_1(N_s)\delta_2(N_s)+9\delta_3(N_s)\right).
\end{equation}
Setting $N_s=1$ in (\ref{eq:c2}) and taking $n_f=3,4,5$, we recover
the values $c_2(1)=0.0372, 0.0609, 0.0838$
obtained in the vacuum \cite{dln}. Moreover,
in (\ref{eq:expandG2}), the sign of successive terms change
as higher order corrections are added to the series. Consequently, it
should be wondered whether this result can drastically be
affected as higher order terms are incorporated to the series
at current energy scales. The highest energy scales reached at the LHC
and measured by the ALICE
and CMS experiments at CERN will provide more reliable comparisons
with our predictions than current experimental studies at RHIC.

\subsubsection{Medium-modified $\boldsymbol{Q_2}$ at NMLLA and 
expansion in $\boldsymbol{{\cal O}(\gamma_0/\sqrt{N_s})}$}

The solution of (\ref{eq:NN2Q}) can also be obtained by setting
$N^{(2)}_q=Q_2N_q^2$ in the equation, using (\ref{eq:solg2})
and taking the MLLA formula for $G_2$ (\ref{eq:expQ2}), one has
\begin{eqnarray}\label{eq:lhsnq}
l.h.s.=2\beta_0\frac{N_c}{C_F}\tilde c_1\gamma_0^4N_Q^2+
\gamma_0^2(Q_2-1)\left(4N_s-\delta_2\sqrt{N_s}\gamma_0+\delta_4(N_s)\gamma_0^2
-4\beta_0r_1\gamma_0^2\right)N_q^2,
\end{eqnarray}
and
\begin{eqnarray}\label{eq:rhsnq}
r.h.s.=\frac{C_F}{N_c}\gamma_0^2\left(N_sG_2
-\frac83\sqrt{N_s}\tilde a_1\gamma_0+
2\left(\frac23\tilde a_1(2\sigma_1+\beta_0)+\tilde a_1c_1+\frac83\tilde a_2(N_s)N_s\right)\gamma_0^2\right)N_g^2.
\end{eqnarray}
After equating (\ref{eq:lhsnq}) and (\ref{eq:rhsnq}) 
we obtain the new exact analytical
solution of (\ref{eq:NN2Q})
\begin{equation}\label{eq:NMLLACORR}
Q_2-1=\frac{N_c}{C_F}\left(\frac{G_2-\tilde\delta_1
\displaystyle{\frac{\gamma_0}{\sqrt{N_s}}}+\tilde\delta_3(N_s)
\displaystyle{\frac{\gamma_0^2}{N_s}}}
{4-\tilde\delta_2\displaystyle{\frac{\gamma_0}{\sqrt{N_s}}}+\tilde\delta_4(N_s)\displaystyle{\frac{\gamma_0^2}{N_s}}}\right)
\frac{r^2}{r_0^2},
\end{equation}
where (see (\ref{eq:nmllaratio}))
$$
\frac{r}{r_0}=1-r_1\frac{\gamma_0}{\sqrt{N_s}}-r_2(N_s)\frac{\gamma_0^2}{N_s}.
$$
Moreover,
\begin{eqnarray}
\tilde\delta_1\!\!&\!\!=\!\!&\!\!\frac83\tilde a_1=2,\quad
\tilde\delta_2=\delta_2,\\
\tilde\delta_3(N_s)\!\!&\!\!=\!\!&\!\!
2\left(\frac23\tilde a_1(2\sigma_1+\beta_0)+\tilde a_1c_1
-\beta_0\tilde c_1+\frac83\tilde a_2(N_s)N_s\right),\\
\tilde\delta_4(N_s)\!\!&\!\!=\!\!&\!\!\delta_4(N_s)-4\beta_0r_1,
\end{eqnarray}
and $G_2$ should be taken from (\ref{eq:solg2}). As before, the size of
NMLLA coefficients $\tilde\delta_3(N_s)$ and $\tilde\delta_4(N_s)$ in the 
vacuum are quite sizable, for $n_f=3$ one has indeed, 
$\tilde\delta_3(1)=3.598$ and $\tilde\delta_4(1)=3.210$, which are close
to $k^2=4$, where $k=2$ labels the rank of the 
second multiplicity correlator.

Performing the same
expansion in $\gamma_0/\sqrt{N_s}$ we obtain the result
\begin{equation}\label{eq:expandQ2}
Q_2-1\approx\frac{N_c}{C_F}\left(\frac13-
\tilde c_1\frac{\gamma_0}{\sqrt{N_s}}+
\tilde c_2(N_s)\frac{\gamma_0^2}{N_s}\right)+{\cal O}\left(\frac{\gamma_0^3}{N_s^{3/2}}\right),
\end{equation}
where the expression for 
$\tilde c_2(N_s)$ follows from (\ref{eq:NMLLACORR}):
\begin{eqnarray}\label{eq:tildec2}
\tilde c_2(N_s)\!\!&\!\!=\!\!&\!\!\frac{1}{12}\left(\frac{\delta_2^2(N_s)}{4}-\tilde\delta_4(N_s)\right)-
\frac{\delta_2(N_s)}{16}
\left(c_1+2+\frac83r_1\right)\\
\!\!&\!\!+\!\!&\!\!\frac14(c_2(N_s)+\tilde\delta_3(N_s))+
\frac{r_1}2(c_1+2)-\frac13(2r_2(N_s)-r_1^2).
\end{eqnarray}
Accordingly, setting $N_s=1$ in (\ref{eq:tildec2}), we find 
the values in the vacuum $\tilde c_2(1)=0.215,0.222,0.229$
respectively for $n_f=3,4,5$ like in \cite{dln}. The sign of successive terms added to the series
(\ref{eq:expandQ2}) shows the wide oscillating property. We give the 
values of $c_2(N_s)$ and $\tilde c_2(N_s)$ in Table \ref{table:c2coeff}.
\begin{table}[htb]
\begin{center}
\begin{tabular}{cccccc}
\hline\hline
$n_f$ & $c_2(N_s)$
& $\tilde c_2(N_s)$ \\ \hline
&&\\
3 & $-0.258-0.016 \sqrt{N_s}+2.505 N_s-2.193 N_s^2$ & 
$-0.168+0.005 \sqrt{N_s}+1.962N_s-1.584 N_s^2$\\
&&\\
4 & $-0.236-0.022 \sqrt{N_s}+2.513 N_s-2.193 N_s^2$ & 
$-0.146+0.007 \sqrt{N_s}+1.946 N_s-1.584 N_s^2$\\
&&\\
5 & $-0.215-0.029 \sqrt{N_s}+2.521 N_s-2.193 N_s^2$ & 
$-0.126+0.009 \sqrt{N_s}+1.930 N_s-1.584 N_s^2$\\
&&\\
\hline\hline
\end{tabular}
\caption{Coefficients $c_2(N_s)$ and $\tilde c_2(N_s)$.}
\label{table:c2coeff}
\end{center}
\end{table}

\subsection{NLO and NNLO results on $\boldsymbol{G_2}$ and $\boldsymbol{Q_2}$}

\begin{figure}[t]
\begin{center}
\epsfig{file=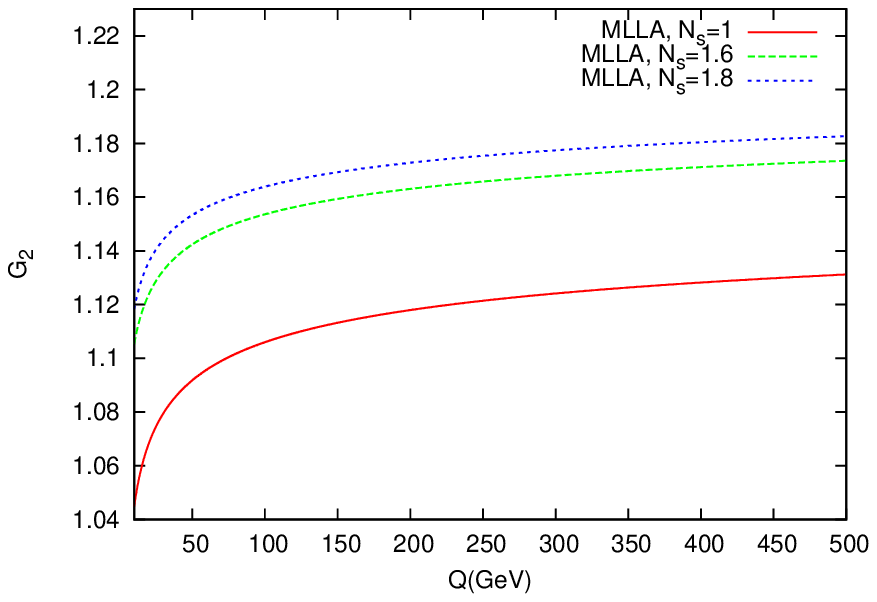,width=7.5truecm}
\qquad
\epsfig{file=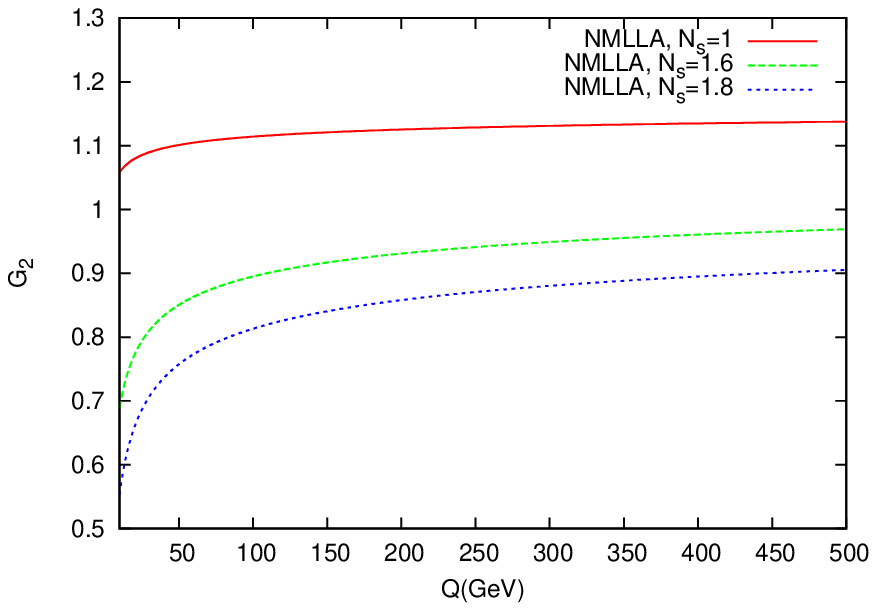,width=7.5truecm}
\caption{MLLA (\ref{eq:expG2}) 
and NMLLA (\ref{eq:expandG2}) second multiplicity correlator inside a gluon jet
as a function of $Q=E\Theta$ in the vacuum ($N_s=1$) and
in the medium ($N_s=1.6$ and $N_s=1.8$).
\label{fig:G2}
}
\end{center}
\end{figure}

\begin{figure}[t]
\begin{center}
\epsfig{file=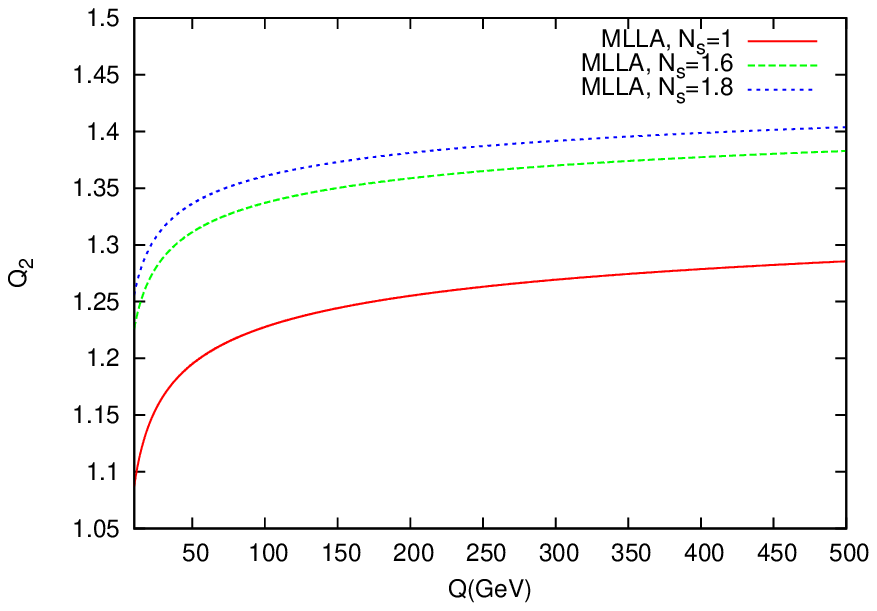,width=7.5truecm}
\qquad
\epsfig{file=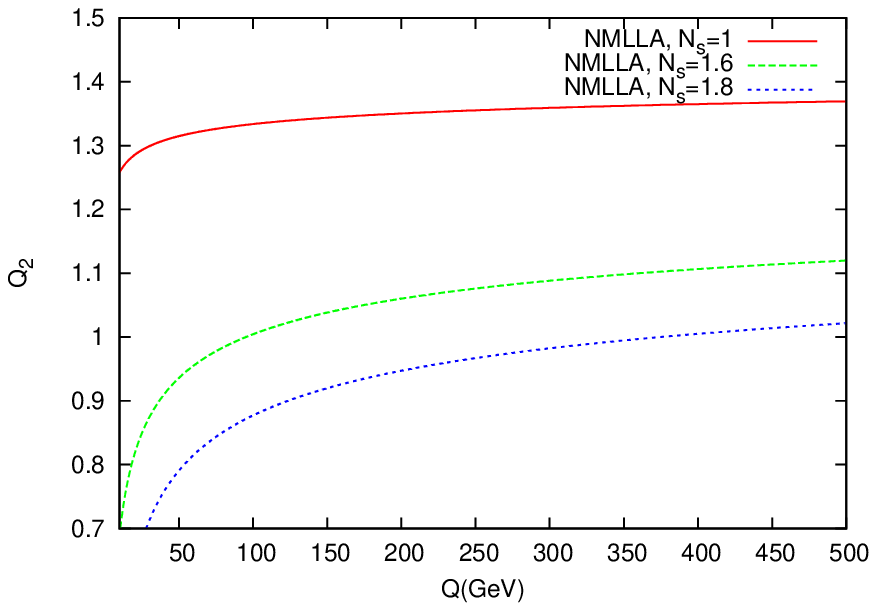,width=7.5truecm}
\caption{MLLA (\ref{eq:expQ2}) 
and NMLLA (\ref{eq:expandQ2}) second multiplicity correlator inside a quark jet
as a function of $Q=E\Theta$ in the vacuum ($N_s=1$) and
in the medium ($N_s=1.6$ and $N_s=1.8$).
\label{fig:Q2}
}
\end{center}
\end{figure}

The MLLA and NMLLA predictions for $G_2(Q)$ (\ref{eq:expandG2}) and $Q_2(Q)$
(\ref{eq:expandQ2}) are depicted respectively in Fig.\,\ref{fig:G2} and  
Fig.\,\ref{fig:Q2}. At MLLA, the second multiplicity correlator increases
as $N_s>1$ and approaches the asymptotic regime $A_2=1+\frac{N_c}{3C_A}$. 
Indeed, as for the MLLA ratio $r(N_s)$ (\ref{eq:mllaratio}), 
the hard corrections ${\cal O}(\gamma_0)$ are suppressed 
by $1/\sqrt{N_s}$, such that the production 
of soft and collinear hadrons
is enhanced, while that of hard collinear
hadrons is more restricted. 
As before, these results provide evidence for the softening of 
jets in the nuclear medium.
However, the NMLLA results 
(\ref{eq:expandG2},\ref{eq:expandQ2}) follow the behavior described in 
section \ref{subsec:figs1} for $r(N_s)$. 
As $N_s>1$, the correlators decrease,
one finds indeed the rough dependence 
$c_2(N_s)\simeq-N_s^2$, $\tilde c_2(N_s)\simeq-N_s^2$ 
(see Table \ref{table:c2coeff}),
which in both cases leads to the unavoidable decrease
of $A_2$ as $N_s$ increases. This result follows 
from the wide oscillating property of the perturbative
series: it is wider in the medium than in the vacuum. That is the reason why,
the more physical MLLA trends can be recovered either by incorporating higher order terms
or by numerically solving the 
evolution equations (\ref{eq:corrg},\ref{eq:corrq}) 
like in \cite{LupiaOchs,sapeta}.

Another interesting feature of these observables concerns the shape of the curves.
They are roughly identical and do not prove to depend on the medium parameter
$N_s$. Moreover, there exists evidence for a flattening of the slopes 
as the jet hardness $Q=E\Theta$ increases for
$N_s\geq1$ (vacuum and medium). This kind of scaling behavior 
is known as the KNO scaling: 
it was discovered by Polyakov in quantum field theory \cite{Poly} and experimentally 
confirmed by $e^+e^-$ 
measurements \cite{experN2} for the second and higher order
multiplicity correlators. This phenomenon implies a jet energy independence of the
normalized multiplicity correlators, which is not affected by
$N_s$ neither at MLLA nor at NMLLA.

\subsection{Role of higher order corrections}

\label{subsec:higher}

In this paragraph we comment on some progresses that could be carried out beyond
the NMLLA approximation. We take the much simpler 
example of the gluon to quark average multiplicity ratio and give the rough 
dependence of the third coefficient $r_3(N_s)$ that can be added to the series 
(\ref{eq:nmllaratio}) in the form
\begin{equation}\label{eq:nnmllaratio}
r=r_0\left(1-r_{1}\frac{\gamma_0}{\sqrt{N_s}}-r_2(N_s)
\frac{\gamma_0^2}{N_s}-r_3(N_s)\frac{\gamma_0^3}{N_s^{3/2}}\right),
\end{equation}
with
$$
r_3(N_s)\stackrel{N_s\gg1}{\propto} N_s^2\left[a_3(N_s)-\tilde a_3(N_s)\right],
$$
where
\begin{equation}
a_3(N_s)\stackrel{N_s\gg1}{\propto}-N_s\zeta(3),\quad
\tilde a_3(N_s)\stackrel{N_s\gg1}{\propto}-\frac{C_F}{N_c}N_s\zeta(3).
\end{equation}
These coefficients follow from (\ref{eq:NGh},\ref{eq:NQh}) by further expanding
the perturbative series and therefore,
\begin{equation}
-r_3(N_s)\stackrel{N_s\gg1}{\propto}N_s^3\left(1-\frac{C_F}{N_c}\right)\zeta(3)>0.
\end{equation}
Thus, replacing $-r_3(N_s)\stackrel{N_s\gg1}{\propto} N_s^3$ in 
(\ref{eq:nnmllaratio}), the third term changes its sign and therefore,
the MLLA trends as $N_s>1$ can be recovered. However, the whole calculation requires the
implementation of the two-loops coupling constant in the solution, and
eventually, the inclusion of the time-like sub-leading splitting functions in the
evolution equations. Nevertheless, as powers of
$N_s$ increase for higher order terms, 
the perturbative approach fails and the
exact numerical solution of the 
evolution equations becomes necessary.

\section{Conclusions}

\label{section:conclusions}

In this paper we have dealt with the medium-modified average multiplicity 
and the medium-modified
second multiplicity correlators in quark and gluon jets. 
Our calculations are based on the 
Borghini-Wiedemann model
\cite{Urs}, which models parton energy loss 
in a dense nuclear medium.
The average multiplicity is found, 
after multiple re-scattering
of the relativistic hard parton in the medium, to be enhanced by the
factor $\sqrt{N_s}$ on the
exponential leading contribution. The former leads, in
particular, to the medium-modified anomalous dimension $\gamma_{\text{med}}$ 
($\gamma\to\gamma_{\text{med}}\approx\sqrt{N_s}\gamma_0$). Corrections 
to the leading double logarithmic contribution of the
average multiplicity arise from both
the MLLA and the NMLLA,
which better account for the energy balance and for the running of 
the coupling constant $\alpha_s$ effects as in the vacuum. 
In particular, the NMLLA average
multiplicity distribution is
softer at NMLLA than at MLLA (see Fig.\,\ref{fig:avemult}), such
that the available phase space for harder collinear
hadronic production becomes restricted. The increase of the average multiplicity
at NNLO is driven by the factor $\propto N_s^{3/2}$ 
(see (\ref{eq:nsmultgnmlla})). 

The MLLA scheme provides a more realistic picture
of the jet quenching through the study of these observables:
such is the case of the medium-modified gluon to quark average multiplicity ratio
$r=N_g/N_q$. Indeed, hard corrections are suppressed
by the extra factor $1/\sqrt{N_s}$, which leads to 
restriction on production
of hard partons in quark and gluon jets. Therefore,
$r$ approaches its asymptotic DLA limit $r_0=N_c/C_F=9/4$ when
the coherent radiation of soft gluons is enhanced by the medium. The 
amplitude of the oscillating series 
turns out to be wider in the medium than in the vacuum
at all energies. 
Nevertheless, the shapes
obtained at MLLA and NMLLA are roughly identical but the series
may require the incorporation of higher order corrections. 
Furthermore, in both approaches, the gluon jets are still more active
than the quark jets in producing secondary particles  but these
characteristics are related to the jet energy dependence
of these observables rather than to the sensitivity to
the parameter $N_s$ in the nuclear medium.

The second multiplicity correlators in quark and gluon jets in the
medium are also computed at MLLA and NMLLA. The multiplicity
fluctuations of individual events must be larger for quark
jets as compared to gluon jets just like in the vacuum.
The MLLA corrections are suppressed by 
$1/\sqrt{N_s}$, such that $A_2$ approaches the
asymptotic DLA regime as $N_s>1$, reproducing the expected physics. 
In addition,
the KNO scaling holds at MLLA and NMLLA in heavy-ion collisions, the 
flattening of the slopes in both the vacuum and the medium is 
roughly reached for the same virtualities $Q>100$ GeV of the
jet energy. As before, the scaling depends on the energy scale
$Q$ rather than on the sensitivity to the nuclear factor $N_s$. 
At NMLLA, the behavior as $N_s>1$ is inverted, but this output can be cured, either
by incorporating higher order terms to the series or by exactly
solving the evolution equations numerically, but this is out of the scope
of this paper.

Finally, our results might lead to more accurate prescriptions for
the behavior of these observables in the presence of the nuclear environment 
if the treatment of parton energy loss
is improved in the future. Furthermore, the
study of parton energy loss and medium-modified observables
would ideally require the re-construction of jets in heavy-ion
collisions. Of course, the huge background makes this task highly delicate.
Nevertheless, thanks in particular to important theoretical
developments on the jet re-constructions algorithms \cite{salam}
in a high-multiplicity environment, future analysis at the LHC 
by ALICE \cite{ALICE} and CMS \cite{CMS} look very promising.

\vskip 3cm

\begin{flushleft}
{\em \underline{Acknowledgments}}: I would like to thank B.A. Kniehl for
supporting my stay at University of Hamburg, as well as S. Albino,
F. Arleo and I. Dremin
for enlightening discussions and useful comments on the manuscript.
\end{flushleft}

\listoffigures

\null



\newpage

\begin{thebibliography}{50}
%
\bibitem{Basics}
Yu.L. Dokshitzer, V.A. Khoze, A.H. Mueller \& S.I. Troyan, Basics of
Perturbative QCD, Editions Fronti\`eres, Paris (1991).

\bibitem{PHENIX}
K. Adcox et al. (PHENIX Collab.), Phys. Rev. Lett. {\bf 88}
(2002) 022301;\newline
S.S. Adler et al. (PHENIX Collab.),
Phys. Rev. Lett. {\bf 91} (2003) 072301.

\bibitem{STAR}
C. Adler et al. (STAR Collab.),
Phys. Rev. Lett {\bf 89} (2002) 202301.

\bibitem{arleo}
F. Arleo, hep-ph/08101193;\\
R. Baier, D. Schiff \& B. G. Zakharov, Ann. Rev. Nucl. Part. Sci. 
{\bf 50} (2000) 37;\\
A. Kovner \& U. A. Wiedemann, in "Quark Gluon Plasma
3", World Scientific, Singapore, hep-ph/0304151;\\
M. Gyulassy, I. Vitev, X.-N. Wang \& B.-W. Zhang,
nucl-th/0302077, ibid.;\\
A. Majumder, J. Phys. G {\bf 34} (2007) S377;\\
For a more recent review, 
see also S. Peign\'e \& A.V. Smilga, hep-ph/08105702.

\bibitem{Urs}
N. Borghini \& U.A. Wiedemann, hep-ph/0506218.

\bibitem{leticia}
N. Armesto, L. Cunqueiro, C. Salgado \& W.C. Xiang,
JHEP {\bf 02} (2008) 048.

\bibitem{sapeta}
S. Sapeta \& U.A. Wiedemann, hep-ph/0809.4251.

\bibitem{wiedemann}
U.A. Wiedemann, Nucl. Phys. B {\bf 588} (2000) 303.

\bibitem{dglap} V.~N. Gribov \& L.~N. Lipatov,
Yad.\ Fiz.\ {\bf15}, 781 (1972) [Sov.\ J. Nucl.\ Phys.\ {\bf15}, (1972) 438];
G. Altarelli \& G. Parisi, Nucl.\ Phys.\ B {\bf 126}, (1977) 298;
Yu.~L. Dokshitser, Zh.\ Eksp.\ Teor.\ Fiz.\ {\bf 73}, (1977) 1216;
[Sov.\ Phys.\ JETP {\bf46}, (1977) 641].

\bibitem{MultTheory}
A.H. Mueller, Nucl. Phys. B {\bf 241} (1984) 141; 
Erratum ibid., B {\bf 241} (1984) 141.

\bibitem{MALAZA}
E.D. Malaza \& B.R. Webber, Phys. Lett. {\bf B} 149 (1984) 501;
E.D. Malaza \& B.R. Webber, Nucl. Phys. {\bf B} 267 (1986) 702.

\bibitem{DREMIN}
I.M. Dremin \& V.A. Nechitailo, Mod. Phys. Lett. A {\bf 9} (1994) 1471;
JETP Lett. {\bf 58} (1993) 945.

\bibitem{dln}
I.M. Dremin, C.S. Lam \& V.A. Nechitailo,
Phys. Rev. D {\bf 61} (2000) 074020.

\bibitem{OPALrgq}
G. Abbiendi et al., [OPAL Collaboration], Phys. Rev. D {\bf 69} (2004) 032002. 

\bibitem{DELPHIrgq}
J. Abdallah et al. [DELPHI Collaboration], Eur. Phys. J. C {\bf 44} (2005) 311.

\bibitem{experN2}
HRS Coll., Phys. Rev. D {\bf 34} (1986) 3304;
AMY Coll., Phys. Rev. D {\bf 42} (1990) 737;
DELPHI Coll., Z. Phys. C-Particles and Fields
{\bf 50} (1991) 185.

\bibitem{DreminGary}
I.M. Dremin \& J.W. Gary, Phys. Rep. {\bf 349} (2001) 301.

\bibitem{TevMult}
D. Acosta et al., Phys. Rev. Lett. {\bf 94} (2005) 171802. 

\bibitem{TevRap}
A.N. Safonov (for CDF Collaboration), Nucl. Phys. B (Proc. suppl.) {\bf 86}
(2000) 55.

\bibitem{PAM} R. Perez-Ramos, F. Arl\'eo \& B. Machet,
Phys. Rev. D {\bf 78} (2008) 014019;
F. Arl\'eo, R. Perez-Ramos \& B. Machet, 
Phys. Rev. Lett. {\bf 100} (2008) 052002. 

\bibitem{termo}
I.M. Dremin, O.S. Shadrin, J. Phys. G {\bf 32} (2006) 963. 

\bibitem{paloma}
N. Armesto, C. Pajares \& P. Quiroga Arias, hep-ph/0809.4428.

\bibitem{LPHD}
Ya.I. Azimov, Yu.L. Dokshitzer, V.A. Khoze \& S.I. Troian,
Z. Phys. C {\bf 27} (1985) 65;
Yu.L. Dokshitzer, V.A. Khoze \& S.I. Troian,
J. Phys. G {\bf 17} (1991) 1585.

\bibitem{perez}
R. Perez-Ramos, hep-ph/0811.2418.

\bibitem{KhozeOchs}
V.A. Khoze \& W. Ochs,
Int. J. Mod. Phys. A {\bf 12} (1997) 2949.

\bibitem{sapetabis}
S. Sapeta \& U.A. Wiedemann, Eur. Phys. J. {\bf C55} (2008) 293.

\bibitem{heinz}
M. Heinz for the STAR Collaboration, nucl-exp/0809.3769.

\bibitem{DokKNO}
Yu.L. Dokshitzer, Phys. Lett. B {\bf 305} (1993) 295.

\bibitem{LupiaOchs}
S. Lupia \& W. Ochs, Phys. Lett. B {\bf 418} (1998) 214.

\bibitem{RPR}
R. Perez-Ramos, JHEP {\bf 06} (2006) 019, and references therein;
R. Perez-Ramos, JHEP {\bf 09} (2006) 014.

\bibitem{Poly}
A.M. Polyakov, Sov. Phys. JETP {\bf 32} (1971) 296.

\bibitem{salam}
M. Cacciari \& G.P. Salam, Phys. Lett. B {\bf 641} (2006) 57.

\bibitem{ALICE}
ALICE collaboration,   
B. Alessandro {\it et al.}, J. Phys. G {\bf 32} (2006) 1295.

\bibitem{CMS}
CMS collaboration,
D. d'Enterria (Ed.) {\it et al.}, J. Phys. G {\bf 34} (2007) 2307.

\end{thebibliography}
\end{document}